\documentclass[a4paper,11pt]{article}
\usepackage{jinstpub} 

\usepackage{caption}
\usepackage{subcaption}
\usepackage{booktabs}
\usepackage{svg}
\usepackage{amsmath}
\usepackage{siunitx}

\title{Particle Hit Clustering and Identification Using Point Set Transformers in Liquid Argon Time Projection Chambers}

\author[a,1]{Edgar E. Robles,\note{Corresponding author.}}
\author[b]{Alejandro Yankelevich,}
\author[b]{Wenjie Wu,}
\author[b]{Jianming Bian,}
\author[a]{and Pierre Baldi}
\affiliation[a]{Department of Computer Science, University of California, Irvine, CA 92697, USA}
\affiliation[b]{Department of Physics and Astronomy, University of California, Irvine, CA 92697, USA}

\emailAdd{roblesee@uci.edu}

\abstract{Liquid argon time projection chambers are often used in neutrino physics and dark-matter searches because of their high spatial resolution. The images generated by these detectors are extremely sparse, as the energy values detected by most of the detector are equal to 0, meaning that despite their high resolution, most of the detector is unused in a particular interaction. Instead of representing all of the empty detections, the interaction is usually stored as a sparse matrix, a list of detection locations paired with their energy values. Traditional machine learning methods that have been applied to particle reconstruction such as convolutional neural networks (CNNs), however, cannot operate over data stored in this way and therefore must have the matrix fully instantiated as a dense matrix. Operating on dense matrices requires a lot of memory and computation time, in contrast to directly operating on the sparse matrix. We propose a machine learning model using a point set neural network that operates over a sparse matrix, greatly improving both processing speed and accuracy over methods that instantiate the dense matrix, as well as over other methods that operate over sparse matrices. Compared to competing state-of-the-art methods, our method improves classification performance by 14\%, segmentation performance by more than 22\%, while taking 80\% less time and using 66\% less memory. Compared to state-of-the-art CNN methods, our method improves classification performance by more than 86\%, segmentation performance by more than 71\%, while reducing runtime by 91\% and reducing memory usage by 61\%.}

\begin{document}
\maketitle
\flushbottom

\section{Introduction}

Experiments in the field of particle physics often create large amounts of data, which is difficult to process at scale by human experts. This data often needs to be manually sorted by these experts, using valuable time that could be used interpreting the data. The advent of high-quality machine learning models has helped automate much of the manual labor required to label these images~\cite{baldi2021science}, but with increased quality, there has also been an increase in computational costs and resources required to run these models. Even large experimental collaborations in the field of particle physics often face strict limits in resource utilization during large-scale simulation and data processing.

The liquid argon time projection chamber (LArTPC) is a common choice of detector technology in neutrino physics and direct dark matter searches due to its very high spatial resolution. The operating principle consists of applying an electric field across a large volume of liquid argon. When charged particles pass through the detector, ionized electrons are accelerated toward the anode end of the drift volume. These drift electrons are usually detected via either a series of wire planes or a grid of charge-detecting pixels. Together with the detection time of the drift electrons, this technology allows for 3D reconstruction of particle trajectories through the detector. These trajectories appear as tracks or showers referred to as "prongs". Particles may also decay in their trajectory, splitting into more particles and creating new prongs. The task at hand is then to perform instance segmentation over these prongs to cluster them as well as to classify each hit into its corresponding particle type for prong identification.

Due to the high spatial resolution, LArTPC images are exceptionally sparse, consisting of an empty background in most of the image except for a few prongs. As such, these are usually represented as sparse matrices, stored as a list of coordinates and values. When performing computations such as the ones used in segmentation machine learning models, these sparse matrices have to be converted into dense matrices, which can take up a lot of resources and slow down training and inference. There have been implementations of differentiable convolution operations on sparse matrices, such as Nvidia's MinkowskiEngine~\cite{choy20194d}. However, the operations need to approximate a convolution in order to save memory. An alternative to using sparse matrices is to represent the sparse image as a point cloud, which only requires coordinates and values to be operated on directly.

Similarly to traditional scintillator cell detectors, LArTPCs with wire-based readout provide multiple 2D views that are subsequently combined to create the 3D reconstruction of particle trajectories. The more novel pixel-based readout for LArTPCs intrinsically provides 3D point cloud representations~\cite{larpix}. However, segmentation over large 3D images can be prohibitively computationally expensive, so images are often reduced to multiple 2D views to save memory. Finally, downsampling is often used to further save on memory when it is necessary to process large volumes, as is the case with events containing long muon tracks.

\subsection{Related work}
The segmentation tasks considered in this work are commonly handled through the Pandora multi-algorithm approach for LArTPC event reconstruction, and a variety of clustering algorithms are available in the Pandora software development kit \cite{pandora, pandora_sdk}. The Wire-Cell software package has also introduced machine-learning based approaches for these tasks \cite{wirecell, wirecell_microboone}, and the PoLAr-MAE model has recently addressed this task with a transformer architecture \cite{polar_mae}. CNNs are often used for event and particle classification at LArTPCs, building on the work of the NOvA CNN \cite{dune_tdr_ii, eventcvn, transformercvn}. Through panoptic segmentation, this work addresses both clustering and particle classification.

We will be interpreting the data as point sets rather than pixels, thus we rely on the Deep Sets~\cite{zaheer2017} framework. This framework has been extended to implement self-attention and graphs in later works. One such work is Point Transformers~\cite{wu2022, wu2024}, A model that implements an attention mechanism between neighboring points in a point cloud. Point Transformer v2 uses k-nearest neighbors to create a graph between points to calculate the attention between closer points, while Point Transformers v3 uses a different serialization technique to save memory usage.

We choose to extend the concepts from point set transformers using Heterogeneous Graph Transformers~\cite{hu2020heterogeneous}, a method that implements attention in heterogeneous graphs. Heterogeneous graphs are graphs where each node is part of a different semantic class, meaning that using different attention weights is able to model the data in a more semantically correct way. This allows us to interpret information from two different views that are related to eachother.

Our model differs from graph attention-based models, such as Graph Attention Transformers~\cite{velickovic2018}, by leveraging the coordinates of each of the points. Rather than just using coordinates of the points as a feature or as inputs to the kNN algorithm that builds the graph, our model uses them to implement a faster pooling algorithm~\cite{simonovsky2017}, which reduces the computation time. We also include a relative positional encoding scheme~\cite{wu2022} in order to decay information from neighbors that are too far away.

\section{Methods}

\subsection{Notation}
Consider a dataset $\mathcal{X}$ of size $N$, where each sample $X^{(i)}$ represents an event from the particle detector. Each event $X^{(i)}$ is split into $M$ views, each view denoted by $X^{(i,j)}$. Each view has a variable number of detections $K^{(i,j)}$. Each detection is described by coordinates $x^{(i,j)}_k \in \mathbb{R}^c$ and values $v^{(i,j)}_k \in \mathbb{R}^d$. Pixel-based TPCs present a homogeneous view as a single 3D point cloud. Wire-based TPCs present heterogeneous views as multiple 2D point clouds. We will treat homogeneous views as a special case of heterogeneous views when explaining the methods for the purpose of easing the burden on notation.

For each pair of points we define an intra-view distance $d_{jj}(x^{(i,j)}_k,x^{(i,j)}_{k'})$ for points within the same view and therefore vector space and an inter-view distance $d_{jj'}(x^{(i,j)}_{k},x^{(i,j')}_{k'})$ for points between different views. Additionally, based on these distances we will define an edge $e^{(i)}_{k,k'} \in \{0, 1\}$ which connects two nodes that may be in the same or different views.

\subsection{Homogeneous attention}
Point attention is calculated by creating a graph between points, using nearest neighbors or other serialization techniques in order to emulate a rolling window, such as the one present in a traditional attention model~\cite{baldi2023quarks}. The attention is then calculated and aggregated over the neighborhood of this graph, for example, if there is a source node $x^{(i,j)}_{k}$ and a destination node $x^{(i,j)}_{k'}$, a query $Q^{(i,j)}_{k}$ is calculated with respect to the source and a key $K^{(i,j)}_{k'}$ and a value a value $V^{(i,j)}_{k'}$ are calculated with respect to the destination for each edge. Each edge within the same view $e^{(i)}_{kk'}$ is then given a score of 
\begin{equation}
w^{(i,j)}_{kk'} = Q_{k}^{(i,j)T}K^{(i,j)}_{k'} + \mathrm{RPE}(x^{(i,j)}_{k}, x^{(i,j)}_{k'}),
\end{equation}
where RPE is a relative positional encoding module, where the difference of the two points are encoded by a linear layer $W$, i.e., 
\begin{equation}
\mathrm{RPE}(x^{(i,j)}_{k}, x^{(i,j)}_{k'}) = W(x^{(i,j)}_{k}-x^{(i,j)}_{k'}).
\end{equation}
The weights are then normalized with a softmax operation to represent the intensities of how much information is required to flow from each edge. This is then used to weigh the value vectors:
\begin{equation}
h^{(i,j)}_{k} = \sum_{k'} \mathrm{softmax}_{\ell}(w^{(i,j)}_{k\ell})_{k'} V^{(i,j)}_{k'},
\end{equation}
leading to a point set attention mechanism.

\subsection{Homogeneous pooling and unpooling}\label{sec:pooling}
A pooling operation is used in U-net-like architectures to create feature representations between points that are further away from each other. The way to extend this concept from CNNs to PSNNs and GNNs is to pool neighbors with each other. This is slow as the nearest neighbors operation is quite expensive, so we can approximate it with a faster method: we first create a grid $G$ of a specific size $g$, then for each square or cube in the grid $G_\ell$, we average out the coordinates and features of the points within that part of the cube, that is,
\begin{equation}
x'_{i,\ell} = \frac{1}{|G_\ell|} \sum_{x_{i,j} \in G_\ell} x_{i,j},
\end{equation}
meaning that the summary of the points in that cell are located in the middle of all the points in that cell, and,
\begin{equation}
v'_{i,\ell} = \frac{1}{|G_\ell|} \sum_{v_{i,j} \in G_\ell} v_{i,j},
\end{equation}
meaning that the features are the average features of all the points in the cell.

Unpooling is done by using the coordinates from a previous step and then broadcasting the pooled point's features into the coordinates that created it. This does create a set where the features will be the same within the grid after it is unpooled, making the residual connections of a U-net vital for the operation to be semantically meaningful.

\subsection{Heterogeneous attention}
Wire-based LArTPCs usually output multiple views of the point clouds, where each view presents a different subset of the spatial dimensions.
This means that the data between different point clouds is related but cannot easily be built into a graph. Using the aforementioned inter-view distance, we are able to build the neighborhood graph. Therefore, for each point we calculate the query $Q^{(i,j'\to j)}_k$, i.e., the query on point $k$ from view $j'$ to view $j$ on sample $i$, and then for each of its neighbors $k'$ we calculate both $K^{(i,j'\to j)}_{k'}$ and $V^{(i,j'\to j)}_{k'}$, that is, the key and values on point $k'$ from view $j'$ to view $j$ on sample $i$. Using these, we  can calculate
\begin{equation}
w^{(i,j'\to j)}_{kk'} = Q_{k}^{(i,j'\to j)T}K^{(i,j'\to j)}_{k'},
\end{equation}
a RPE cannot be used here due to both samples being defined on different spaces, making them hard to compare. This weight is then normalized using a softmax operation over its neighbors and then used in a weighted sum to calculate the output of the attention module,
\begin{equation}
{h'}^{(i,j)}_{k} = \sum_{k'} \mathrm{softmax}_{\ell}(w^{(i,j'\to j)}_{k\ell})_{k'} V^{(i,j'\to j)}_{k'}.
\end{equation}

\subsection{Heterogeneous pooling}
When dealing with multiple views from the same detector, the views may be defined in completely different vector spaces, so while we may be able to compare distances to determine nearest neighbors or grids to pool points together, heterogeneous points cannot be pooled together. Therefore, we treat each view separately and pool using a grid pool. Pooling is then done per view, using a voxel pooling method~\cite{simonovsky2017}, in the same manner as with homogeneous pooling described in section~\ref{sec:pooling}, creating a grid and then averaging out the values of all the points within each point of the grid, and positioning the point in the mean of all the points within the created voxel. 

Unpooling is performed using skip connections, the points are upsampled to the same coordinates that they were previously pooled from, only using information from the same view. 

\subsection{Architecture}
\begin{figure*}
    \centering
    \includegraphics[width=0.95\textwidth]{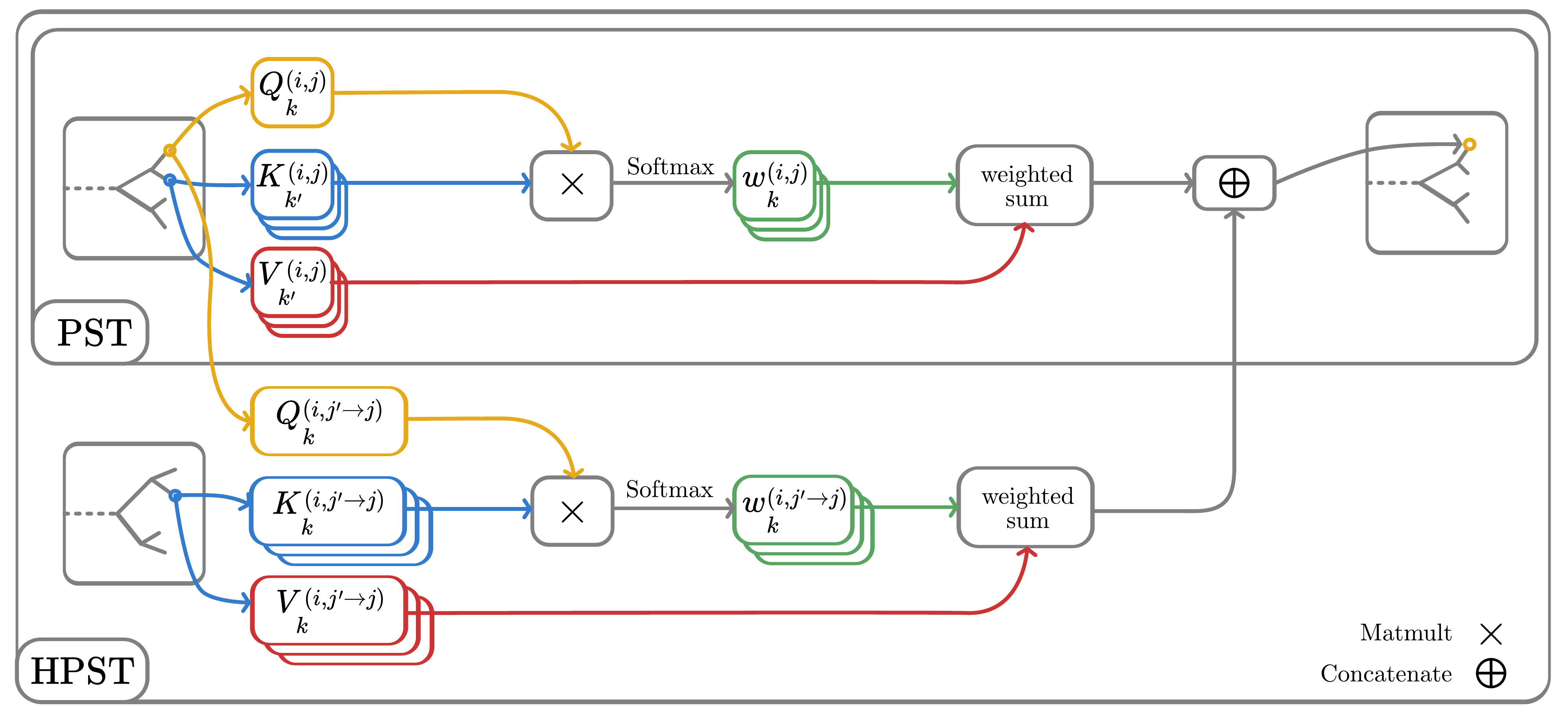}
    \caption{Block diagram of the attention mechanism. The top path describes the intra-view attention mechanism, and the bottom path describes the inter-view mechanism. The top section labeled PST is the attention mechanism used in the point set transformer, while HPST uses both the top and bottom sections.}\label{fig:attn}
\end{figure*}
\begin{figure*}
    \centering
    \includegraphics[width=0.95\textwidth]{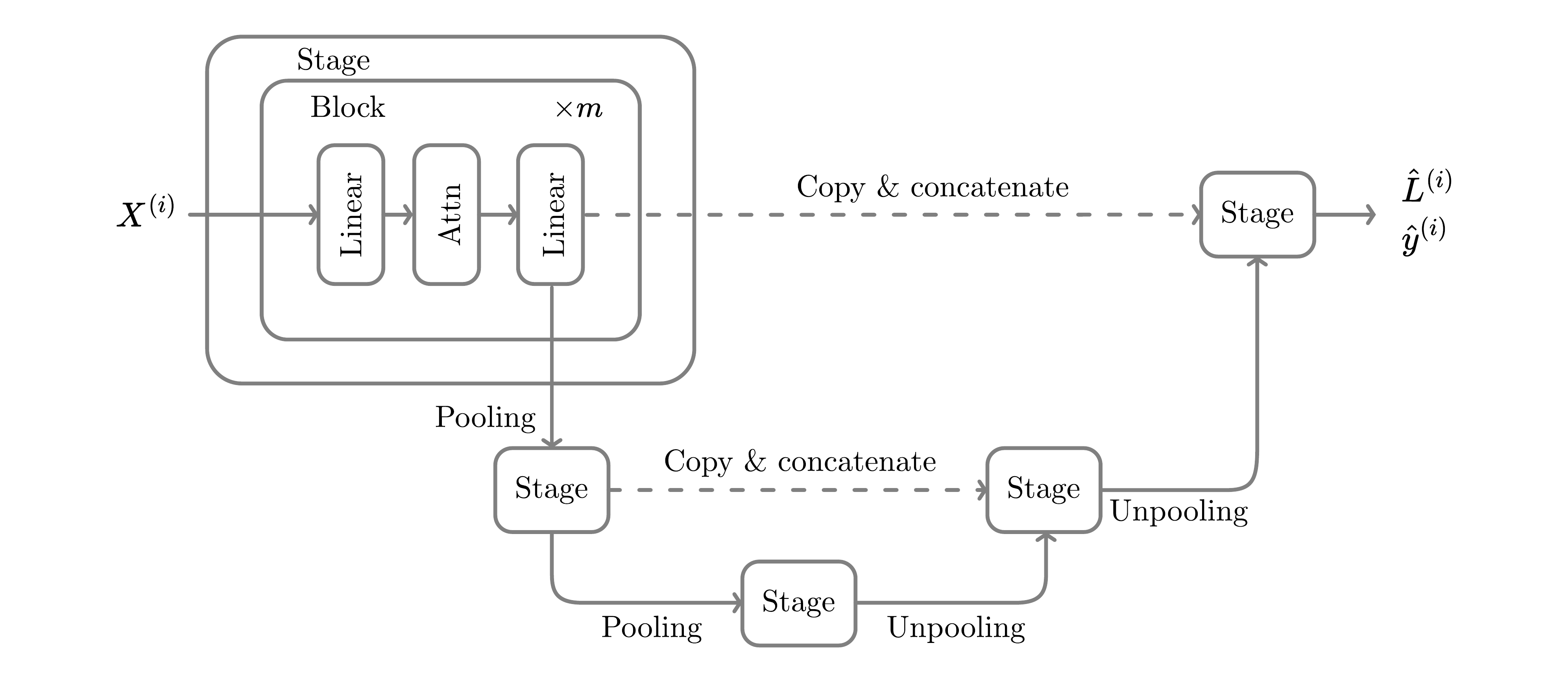}
    \caption{Architecture of the neural network. The attention block is described in Figure~\ref{fig:attn}. The number of stages can be arbitrarily increased by adding stages to both the pooling and unpooling sides.}\label{fig:arch}
\end{figure*}
The network is structured like a U-net~\cite{Ronneberger2015Unet}, where attention layers act as the convolutions and the grid pooling and unpooling function as the pooling method. The architecture is described in Figure ~\ref{fig:arch}. The UNet is divided into $2n$ stages. Each stage contains $m$ blocks, where each block has an attention module. Following each stage in the first $n$ stages is a pooling step, which reduces the number of points. The next $n$ stages are followed by an unpooling stage, which uses the coordinates of the points of a previous block, as well as concatenates the features of the previous block. The dimensionality of the embeddings is doubled at each stage during the first half and halved at each stage during the second half. Intra-view attention is calculated on each stage in order to ensure that the information mixing between views is done locally (in the earlier stages) and globally (in the later stages). 

\subsection{Loss function}
The network performs two tasks simultaneously: instance segmentation, selecting separate prongs from each other; and semantic segmentation, classifying each detection into a particle type. As such, the loss function used is separated into two parts,
\begin{equation}
\mathcal{L} = \lambda\mathcal{L}_{\mathrm{sem}} + (1-\lambda)\mathcal{L}_{\mathrm{ins}}.
\end{equation}
Semantic segmentation is a simple classification problem, so we use multi-class cross-entropy to calculate this loss:
\begin{align}
    \begin{split}
\mathcal{L}_{\mathrm{sem}} & = \sum_{X^{(i)} \in \mathcal{X}} \sum_{X^{(i)} \in x^{(i,j)}_{k}} \\
& \mathrm{CE}\left(\mathrm{softmax}_{k'}\left(f(X_i)^{(i,j)}_{k'}\right),y^{(i,j)}_{k}\right),
    \end{split}
\end{align}
where $y^{(i,j)}_{k}$ is the correct semantic label of the detection. 

Instance segmentation is done by minimizing the loss calculated by the best assignment between the predicted labels and the real labels. If point $x^{(i,j)}_{k}$ belongs to the segment $L^{(i,j)}_{k}$, then the loss calculated is
\begin{align}
\begin{split}
\mathcal{L}_{\mathrm{ins}} & = \sum_{X^{(i)} \in \mathcal{X}} \min_{\phi \in \Sigma} \sum_{x^{(i,j)}_{k} \in X^{(i)}} \\
& \mathrm{CE}\left(\mathrm{softmax}_{k'}\left(f(X_i)^{(i,j)}_{k'}\right),\phi\left(L^{(i,j)}_{k}\right)\right),
\end{split}
\end{align}
where $\Sigma$ is the set of all permutations of labels, allowing a unique assignment of one label to another. The optimal assignment of the labels is solved using a linear sum assignment solver~\cite{crouse2016assignment}. The linear sum assignment solver, also known as the Hungarian algorithm is a method to assign a bipartite graph maximizing a quantity in polynomial time. This allows us to calculate the loss function without needing to check every possible assignment combination. This is a standard method used to train object segmentation models and does not affect the inference time, only the training time, and only scales polynomially with respect to the number of possible object segments in the model, picked as a hyperparameter.

\section{Experiments and Results}
\subsection{Dataset}
We consider here a LArTPC with square \SI{5}{mm} pixel-based readout. The TPC is \SI{2}{m} x \SI{2}{m} x \SI{7}{m}  in $x,y,z$ with a \SI{2}{m} drift length along x. $\nu_e$ and $\nu_\mu$ at energies are simulated with GENIE \cite{Andreopoulos:2015wxa} in the $+z$ direction with uniform neutrino energy up to \SI{10}{GeV}. The energy deposition in liquid argon is then simulated with GEANT4 \cite{Geant:2017ats}. The dataset consists of 100,000 $\nu_e$ and $\nu_\mu$ events each with 74\% of events interacting through the charged current and the rest through neutral current. Additionally, we created another dataset where the odd pixels of the Z dimension were assigned to the XZ view by removing the Y coordinate and the even pixels were assigned to the YZ view by removing the Y coordinate in order to simulate similar multi-view images to those produced by wire LArTPCs.

\subsection{Performance evaluation}
Six models were trained and evaluated, two graph attention network (GAT)~\cite{velickovic2018} based models, one for the multi-view case and one for the single-view case, a 2D CNN-based model (R-CNN)~\cite{girshick2015}, only used for the single view case, a heterogeneous point set transformer for the multi view case, a and point set transformer for the multi-view case and single-view case. We performed a hyperparameter sweep over the number of layers, the layer size, the number of neighbors to use in the nearest neighbors calculation, and the learning rate, sampling 60 random hyperparameters in the grid. The hyperparameters picked for the three networks were the number of neighbor connections (4, 8), the number of stages of the neural network (2, 3, 4), the size of the embeddings inside the neural network (128, 256, 512) and the learning rate (between 1e-4 and 1e-1). The range of the parameters was chosen according to the memory restrictions of our targeted production environment. The same ranges were used for all 3 of them as they all had comparable parameters. The training and testing was done on a server using an Intel(R) Xeon(R) CPU E5-2640 v4 @ 2.40GHz, 503G of RAM and 4xNVIDIA Titan V. 

The hyperparameter sweep was performed over one learning rate cycle with a cosine annealing scheduler, over 64 epochs, using 10\% of the dataset. The model with the best accuracy on the segmentation's class labels in the validation set was selected as the one with the best hyperparameters. The resulting models with the best hyperparameters were trained for 4 learning rate cycles, each cycle being 64 epochs long for a total of 256 epochs using an AdamW optimizer. The results of this optimization for the 3D PST can be seen in table~\ref{tbl:hyperparamters}

\begin{table}[]
\centering
\caption{Best hyperparameters for the 3D PST}
\label{tbl:hyperparamters}
\begin{tabular}{ll}
\toprule
Hyperparameter    & Value     \\
\midrule
Learning Rate     & 0.0006323 \\
Number of stages  & 3         \\
Embed size        & 256       \\
Neighbors         & 8         \\
Initial grid size & 8         \\
\bottomrule
\end{tabular}
\end{table}

\subsubsection{Classification and Segmentation accuracy}
We evaluate the accuracy of the classification and instance segmentation of each point for each model. The results can be seen in table~\ref{tbl:speedmemory}. As we can see, we gain an advantage over using a traditional GAT with a more efficient implementation of attention, as well as using pooling to our advantage, as evidenced by the jump in performance between the 3D GAT and our 3D PST.

Using all the information available in 3D images also helps increase accuracy. Matching prongs between views is an especially hard task, so 3D images will inherently have better performance for segmentation accuracy, as they are single view models. HPST is able to bridge the gap by sharing data between the views, and it is able to improve the performance over the 2D GAT and R-CNN by using less parameters than the 2D PST, this creates a tradeoff where the view sharing can give you good performance with a smaller model, while the PST can have higher accuracy due to being able to reach more neighbors within that same view.

\subsubsection{Efficiency and purity of segmentation}
\begin{figure*}
    \centering
    \includegraphics[width=0.95\textwidth]{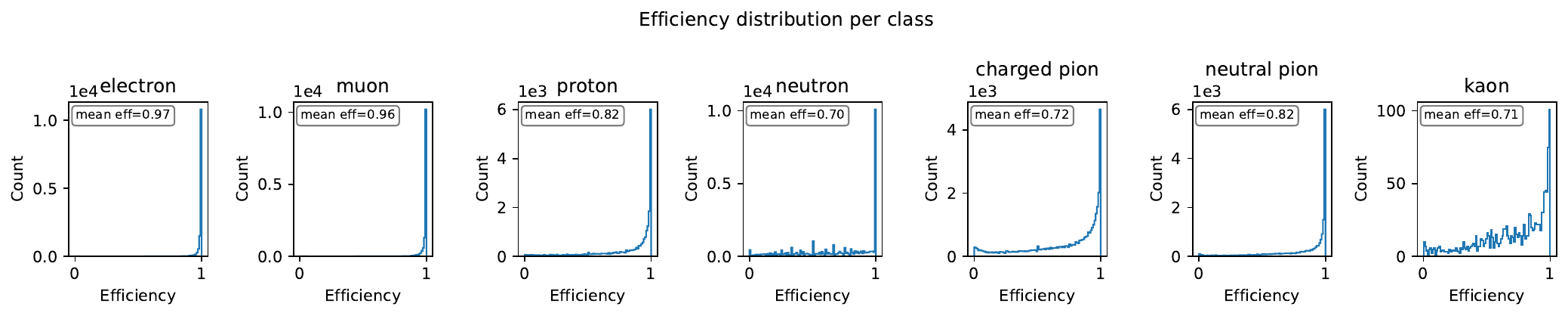}
    \includegraphics[width=0.95\textwidth]{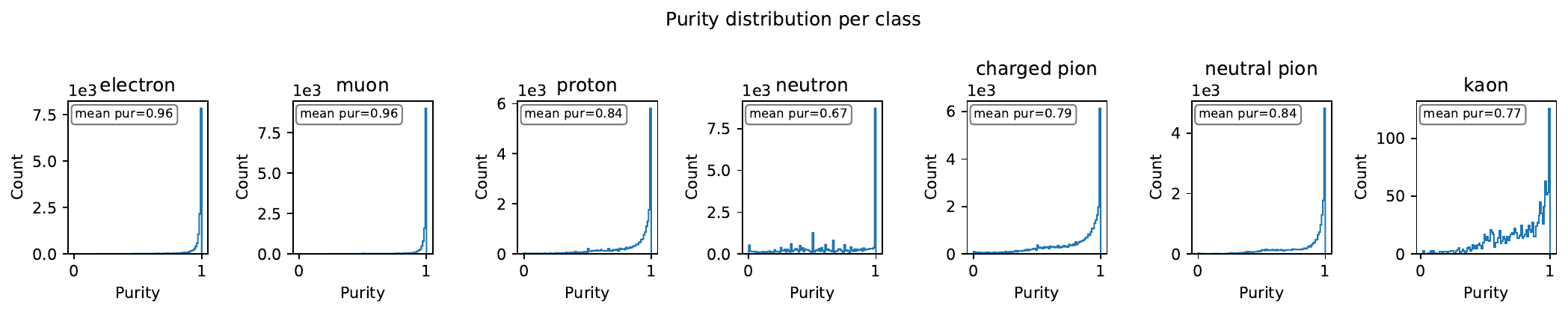}
    \caption{Distribution of prong efficiency and purity}\label{fig:dist}
\end{figure*}
For each prong, we calculated the efficiency and purity of the classification, allowing for multiple predicted prongs to be assigned to a single prong. Efficiency is defined as the percentage of a predicted prong that is assigned to the correct prong. Purity is the percentage of the true prong that is predicted correctly. These are metrics used in particle physics that should be balanced, as raising the purity can often lower the efficiency and vice versa. In figure~\ref{fig:dist} we can see the distribution of the purity and efficiency in each prong. As we can see, the segmentation results are generally good, especially in the majority classes (muons and electrons).

\subsubsection{Speed and memory usage}
\begin{table*}
  \caption{Speed and memory usage for each model compared to their performance.}
  \label{tbl:speedmemory}
  \centering
  \begin{tabular}{lllll}
    \toprule
    Model     & Memory & Time per   & Classification & Instance segmentation\\
    & usage (MiB) & sample (s) &  OVR AUC & accuracy\\
    \midrule
    2D R-CNN        & $440.5 \pm 51.04$  & $1.5752 \pm 0.091$    & 0.526     & 0.518 \\
    2D GAT          & $\mathbf{88.6 \pm 7.56}$   & $\mathbf{0.2300 \pm 0.025}$    & 0.833     & 0.659 \\
    2D HPST (ours)  & $99.1 \pm 7.39$   & $0.3542 \pm 0.019$    & 0.936     & 0.779 \\
    2D PST (ours)   & $138.1 \pm 11.29$   & $0.2539 \pm 0.021$    & \textbf{0.949}     & \textbf{0.827} \\
    \midrule
    3D GAT          & $506.1 \pm 30.13$  & $0.7216 \pm 0.060$    & 0.859     & 0.727 \\
    3D PST (ours)   & $\mathbf{170.2 \pm 9.65}$  & $\mathbf{0.1401 \pm 0.012}$    & \textbf{0.982}     & \textbf{0.889} \\
    \bottomrule
  \end{tabular}
\end{table*}
We benchmarked the three models by running inference on 100 samples, with a batch size of 1, measuring the peak memory increase between the start of inference and the end of inference, in order to remove as much overhead as possible. We evaluated the time it takes for these 100 inferences and the memory used in each of them. We additionally used a Fast R-CNN~\cite{girshick2015} as a comparison in order to evaluate how much memory is saved by evaluating the data as a point cloud. The results can be seen in table~\ref{tbl:speedmemory}. As we can see, memory usage is greatly decreased when comparing a regular CNN model to the sparse methods like graph neural networks and point set neural networks, even when projecting a 3D voxel into two 2D views. 

Although 2D models are able to maintain a lower memory usage profile due to merging obscured points and removing at least 1/3 of the data, our 3D model presents a significant increase in performance, especially when comparing the segmentation accuracy to the 2D models. Our 3D model has a significant enough increase in accuracy to justify the increase in memory usage when compared to the models that do not use all three dimensions. Although the increase in memory usage is significant, the memory usage is still within the memory usage required by the environment in which it will be deployed.

\subsection{Qualitative evaluation}
\begin{figure*}
    \centering
    \includegraphics[width=0.95\textwidth]{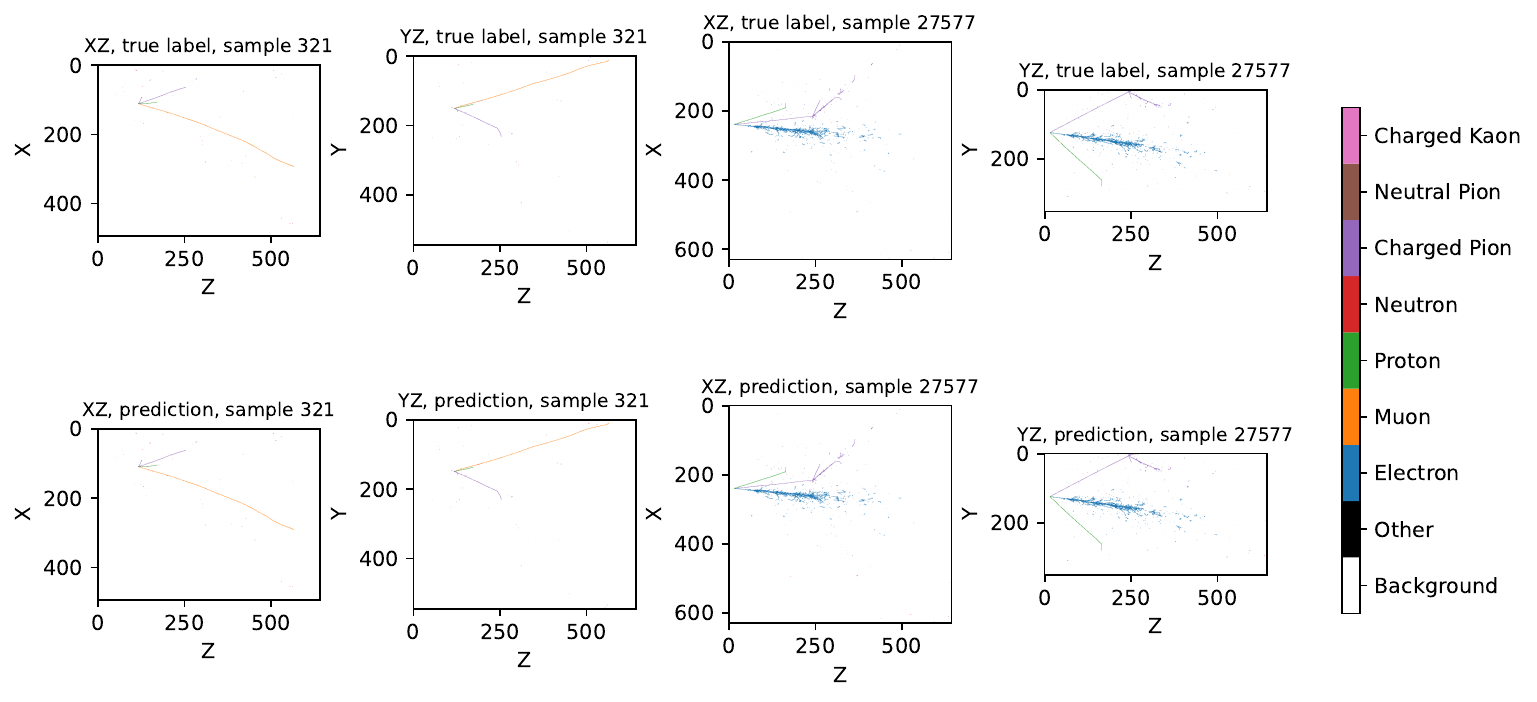}
    \caption{Two example events from the test set. The left two columns show the X and Y views of a muon neutrino event with a long muon track (blue), and the right two show an electron neutrino event with a prominent electron shower (black). The top row shows each hit's true particle label and the bottom row shows the network's predicted segmentation each colored according to the particle class that had the majority of hits classified as such in the segment.}\label{fig:plots}
\end{figure*}

\begin{figure*}
    \centering
    \includegraphics[width=0.95\textwidth]{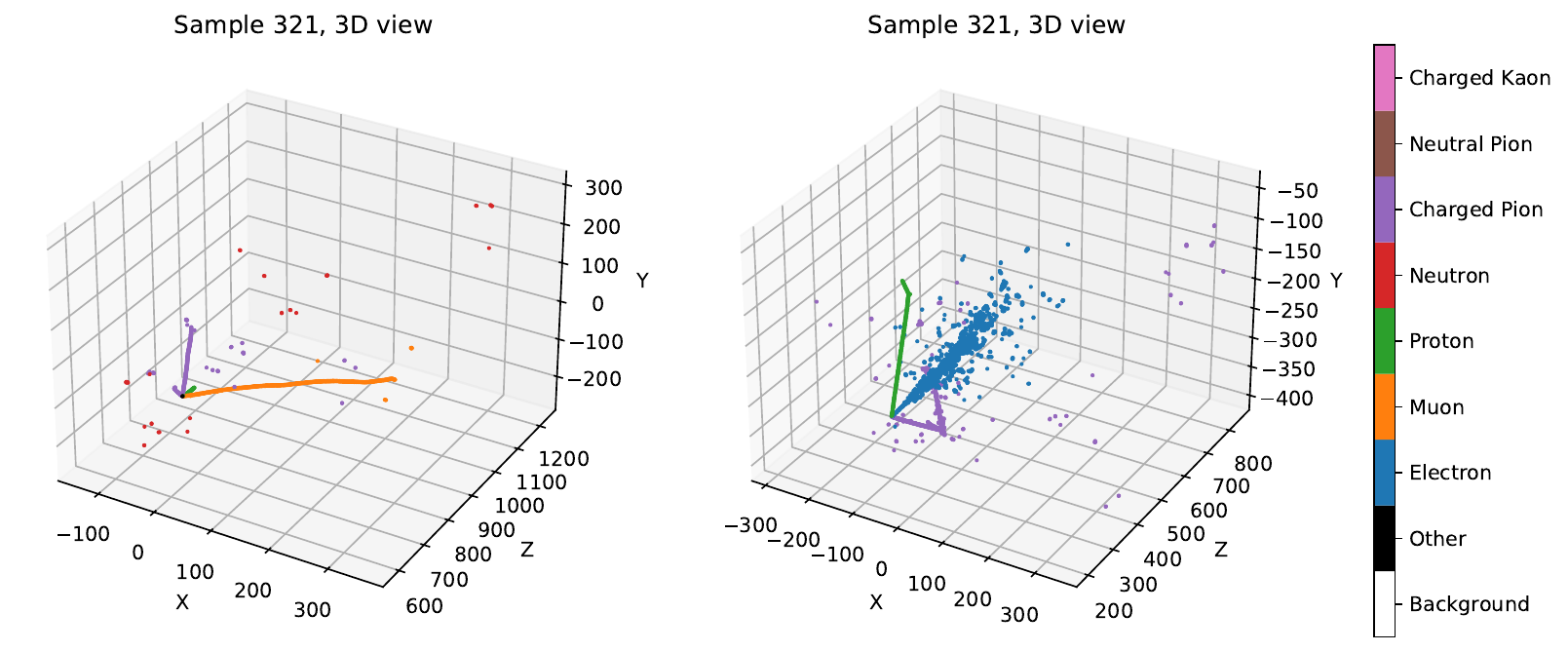}
    \caption{The two example events from figure~\ref{fig:plots} from the test set visualized as 3D plots.}\label{fig:plot3d}
\end{figure*}

Figures~\ref{fig:plots} and~\ref{fig:plot3d} show two samples from the test set. They are colored according to the most common particle predicted by the model in the segmentation. The muon produced by a numu charged current interaction event leaves a long track. Protons and charged pions also visible in this event leave tracks as well, but such tracks are generally shorter. The nue event produces a prominent electron shower. Separation between tracks and showers is an easier task compared to identifying particles with similar topologies, especially in the case of protons vs pions. These particles make up the hadronic portions of neutrino interactions, which are less understood compared to the leptonic portions from electrons and muons.

\section{Limitations and Conclusions}
While the claims of memory efficiency will generally hold, although for different datasets this might not be the case. The representation of a sparse matrix is more efficient than a dense matrix until a certain point, where the storage of the coordinates becomes bigger than just storing a dense matrix. Point set operations can also greatly increase in complexity as the number of points grows, resulting in a much slower algorithm. However, these are not the regimes found in data produced by neutrino detectors. 

Improvements can also be made to the attention mechanism. The current implementation calculates attention manually, instead of using the more optimized FlashAttention~\cite{tri2022flash}, meaning that memory usage and speed can be further reduced. FlashAttention cannot directly be used since it is limited to fixed length sequences, however, a similar strategy could be implemented to speed up point transformer operations. Using nearest neighbors to encode the connections between points is also not necessarily the most efficient method for this dataset. Point Transformers v3 (PTv3)~\cite{wu2024} demonstrates that point transformers can achieve the same performance using fewer connections than a graph based on nearest neighbors. In the PTv3 paper, this is achieved by serializing the points with a space-filling curve, drastically reducing the memory usage for one of the most expensive operations in the model's calculations. 

In general, point set transformers perform very well compared to GNNs and CNNs in this task. PSTs strike a balance between memory usage, time, and performance that makes them a great fit for this application.

\bibliographystyle{plain}
\bibliography{bib}

\begin{thebibliography}{10}

\bibitem{pandora}
R.~Acciarri et~al.
\newblock The {Pandora} multi-algorithm approach to automated pattern recognition of cosmic-ray muon and neutrino events in the {MicroBooN}e detector.
\newblock {\em The European Physical Journal C}, 78(1):82, Jan 2018.

\bibitem{Andreopoulos:2015wxa}
Costas Andreopoulos, Christopher Barry, Steve Dytman, Hugh Gallagher, Tomasz Golan, Robert Hatcher, Gabriel Perdue, and Julia Yarba.
\newblock {The GENIE Neutrino Monte Carlo Generator: Physics and User Manual}, 10 2015.

\bibitem{eventcvn}
A.~Aurisano, A.~Radovic, D.~Rocco, A.~Himmel, M.~D. Messier, E.~Niner, G.~Pawloski, F.~Psihas, A.~Sousa, and P.~Vahle.
\newblock {A Convolutional Neural Network Neutrino Event Classifier}.
\newblock {\em JINST}, 11(09):P09001, 2016.

\bibitem{baldi2021science}
Pierre Baldi.
\newblock {\em Deep Learning in Science}.
\newblock Cambridge University Press, 2021.

\bibitem{baldi2023quarks}
Pierre Baldi and Roman Vershynin.
\newblock The quarks of attention: Structure and capacity of neural attention building blocks.
\newblock {\em Artificial Intelligence}, 319:103901, 2023.
\newblock Also: arXiv:2202.08371.

\bibitem{choy20194d}
Christopher Choy, JunYoung Gwak, and Silvio Savarese.
\newblock 4d spatio-temporal convnets: Minkowski convolutional neural networks.
\newblock In {\em Proceedings of the IEEE Conference on Computer Vision and Pattern Recognition}, pages 3075--3084, 2019.

\bibitem{crouse2016assignment}
David~F. Crouse.
\newblock On implementing 2d rectangular assignment algorithms.
\newblock {\em IEEE Transactions on Aerospace and Electronic Systems}, 52(4):1679--1696, 2016.

\bibitem{tri2022flash}
Tri Dao, Dan Fu, Stefano Ermon, Atri Rudra, and Christopher R\'{e}.
\newblock Flashattention: Fast and memory-efficient exact attention with io-awareness.
\newblock In S.~Koyejo, S.~Mohamed, A.~Agarwal, D.~Belgrave, K.~Cho, and A.~Oh, editors, {\em Advances in Neural Information Processing Systems}, volume~35, pages 16344--16359. Curran Associates, Inc., 2022.

\bibitem{larpix}
D.A. Dwyer, M.~Garcia-Sciveres, D.~Gnani, C.~Grace, S.~Kohn, M.~Kramer, A.~Krieger, C.J. Lin, K.B. Luk, P.~Madigan, C.~Marshall, H.~Steiner, and T.~Stezelberger.
\newblock Larpix: demonstration of low-power 3d pixelated charge readout for liquid argon time projection chambers.
\newblock {\em Journal of Instrumentation}, 13(10):P10007, oct 2018.

\bibitem{Geant:2017ats}
{Geant4 Collaboration}.
\newblock Geant4 10.4 release notes.
\newblock {\em geant4-data.web.cern.ch, https://geant4-data.web.cern.ch/ ReleaseNotes/ReleaseNotes4.10.4.html}, 2017.

\bibitem{girshick2015}
Ross Girshick.
\newblock Fast r-cnn.
\newblock In {\em 2015 IEEE International Conference on Computer Vision (ICCV)}, pages 1440--1448, 2015.

\bibitem{hu2020heterogeneous}
Ziniu Hu, Yuxiao Dong, Kuansan Wang, and Yizhou Sun.
\newblock Heterogeneous graph transformer.
\newblock In {\em Proceedings of The Web Conference 2020}, WWW '20, page 2704–2710, New York, NY, USA, 2020. Association for Computing Machinery.

\bibitem{pandora_sdk}
J.~S. Marshall and M.~A. Thomson.
\newblock The {Pandora} software development kit for pattern recognition.
\newblock {\em The European Physical Journal C}, 75(9):439, Sep 2015.

\bibitem{Ronneberger2015Unet}
Olaf Ronneberger, Philipp Fischer, and Thomas Brox.
\newblock U-net: Convolutional networks for biomedical image segmentation.
\newblock In Nassir Navab, Joachim Hornegger, William~M. Wells, and Alejandro~F. Frangi, editors, {\em Medical Image Computing and Computer-Assisted Intervention -- MICCAI 2015}, pages 234--241, Cham, 2015. Springer International Publishing.

\bibitem{transformercvn}
Alexander Shmakov, Alejandro~J Yankelevich, Jianming Bian, and Pierre Baldi.
\newblock Interpretable joint event-particle reconstruction at {NOvA} with sparse cnns and transformers.
\newblock In {\em Machine Learning and the Physical Sciences, NeurIPS}, 2023.

\bibitem{simonovsky2017}
Martin Simonovsky and Nikos Komodakis.
\newblock Dynamic edge-conditioned filters in convolutional neural networks on graphs.
\newblock In {\em 2017 IEEE Conference on Computer Vision and Pattern Recognition (CVPR)}, pages 29--38, 2017.

\bibitem{dune_tdr_ii}
{The DUNE Collaboration}.
\newblock Deep underground neutrino experiment ({DUNE}), far detector technical design report, volume ii: {DUNE} physics, 2020.

\bibitem{wirecell_microboone}
{The MicroBooNE collaboration}.
\newblock Neutrino event selection in the {MicroBooNE} liquid argon time projection chamber using {Wire-Cell 3D} imaging, clustering, and charge-light matching.
\newblock {\em Journal of Instrumentation}, 16(06):P06043, jun 2021.

\bibitem{velickovic2018}
Petar Veličković, Guillem Cucurull, Arantxa Casanova, Adriana Romero, Pietro Liò, and Yoshua Bengio.
\newblock Graph attention networks, 2018.

\bibitem{wu2024}
Xiaoyang Wu, Li~Jiang, Peng-Shuai Wang, Zhijian Liu, Xihui Liu, Yu~Qiao, Wanli Ouyang, Tong He, and Hengshuang Zhao.
\newblock Point transformer v3: Simpler, faster, stronger.
\newblock In {\em CVPR}, 2024.

\bibitem{wu2022}
Xiaoyang Wu, Yixing Lao, Li~Jiang, Xihui Liu, and Hengshuang Zhao.
\newblock Point transformer v2: Grouped vector attention and partition-based pooling.
\newblock In {\em NeurIPS}, 2022.

\bibitem{polar_mae}
Sam Young, Yeon jae Jwa, and Kazuhiro Terao.
\newblock Particle trajectory representation learning with masked point modeling, 2025.

\bibitem{wirecell}
H.W. Yu, M.~Bishai, W.Q. Gu, M.F. Lin, X.~Qian, Y.H. Ren, A.~Scarpelli, B.~Viren, H.Y. Wei, H.Z. Yu, K.~Yu, and C.~Zhang.
\newblock Augmented signal processing in liquid argon time projection chambers with a deep neural network.
\newblock {\em Journal of Instrumentation}, 16(01):P01036, jan 2021.

\bibitem{zaheer2017}
Manzil Zaheer, Satwik Kottur, Siamak Ravanbakhsh, Barnabas Poczos, Russ~R Salakhutdinov, and Alexander~J Smola.
\newblock Deep sets.
\newblock In I.~Guyon, U.~Von Luxburg, S.~Bengio, H.~Wallach, R.~Fergus, S.~Vishwanathan, and R.~Garnett, editors, {\em Advances in Neural Information Processing Systems}, volume~30. Curran Associates, Inc., 2017.

\end{thebibliography}
\end{document}